\documentstyle[seceq,twoside,epsf,subfigure]{ptptex}
\notypesetlogo  
\title{Light $\sigma$-Meson Production 
in Excited $\Upsilon$ Decay Processes I\\
  --- Analyses --- }
\author{%
Toshihiko {\sc Komada}, Shin {\sc Ishida} and Muneyuki {\sc Ishida}$^{*}$  }
\inst{%
 Atomic Energy Research Institute, 
College of Science and Technology\\
Nihon University, Tokyo 101-0062, Japan\\
$^{*}$Department of Physics, Tokyo Institute of Technology\\
Tokyo 152-8551, Japan
}
\abst{%
We analyze 
the $\pi\pi$ production amplitudes in the excited $\Upsilon$
decay processes, 
$\Upsilon (2S)\to\Upsilon (1S)\pi^+\pi^-$,  $\Upsilon (3S)\to\Upsilon (1S)\pi^+\pi^-$ and
$\Upsilon (3S)\to\Upsilon (2S)\pi^+\pi^-$,
and the $\pi\pi$ and $K\bar K$ production amplitudes 
in the charmonium decay processes, 
$\psi (2S)\to J/\psi \pi^+\pi^-$ and 
$J/\psi\to \phi \pi^+\pi^- ,\ \phi K^+K^-$.
The amplitudes are parametrized by the sum of 
Breit-Wigner amplitudes for the $\sigma$ and the other relevant particles and 
of the direct $2\pi$-production amplitude,
following the VMW method. 
All the $\pi\pi$ (and $K\bar K$) invariant mass spectra are 
reproduced well with the $\sigma$ Breit-Wigner amplitude by using
the common parameters,
$m_\sigma =526^{+48}_{-37} $MeV and 
$\Gamma_\sigma =301^{+145}_{-100} $MeV,
which is almost consistent with the values obtained in our previous 
phase shift analyses.
This strongly suggests the existence of light $\sigma$-meson.
}

\begin{document}
\maketitle

\setcounter{tocdepth}{4}

\section{Introduction}
The $\sigma$-meson is a chiral partner of $\pi$-meson, 
and it is theoretically
important from the viewpoint of chiral symmetry breaking. 
However, its existence
had been neglected phenomenologically, mainly being based on 
the negative result of the
analyses of $I=0$ $S$-wave $\pi\pi$ scattering phase shift. 

In most of the $\pi\pi$-production experiments, 
a large event concentration or a bump structure 
in the spectra of $\pi\pi$ invariant mass 
$m_{\pi\pi}$ around 500 MeV had been observed, 
however, conventionally it was not regarded as
$\sigma$-resonance, but as a mere $\pi\pi$-background, under influence of
the so callled ``universality argument."\cite{rf1} 
In this argument, it is stated that 
because of the unitarity of $S$-matrix and of the analyticity of the amplitudes,
the $\pi\pi$ production amplitude ${\cal F}$ takes the form
${\cal F}=\alpha (s){\cal T}$ (${\cal T}$ being $\pi\pi$ scattering 
amplitude),
with a slowly varying real function $\alpha (s)$.
The pole position of $S$-matrix is determined solely
through the analysis of ${\cal T}$, which was believed to have 
no light $\sigma$-pole at that time.

Recently the data of $\pi\pi$-scattering phase shift 
have been reanalyzed by many groups\cite{rf2}
including ours\cite{rf3} and the existence of light $\sigma (400\sim 700)$ is 
strongly suggested.
The result of no $\sigma$-existence in the conventional analyses is 
pointed out\cite{rf4} to be due to the lack of consideration on the
cancellation mechanism guaranteed by chiral symmetry, 
and shown to be not correct. 
Furthermore, we have pointed out that 
the ``universality argument" should be revised, taking into account 
the quark physical picture.\cite{rf5} 
Actually the ``effective" $\alpha (s)(\equiv {\cal F}/{\cal T})$, 
thus corrected,
is not a slowly varying function but rapidly varying function 
even with real poles.

Now the many $\pi\pi$-production experiments must be reanalyzed
independently from ${\cal T}$ 
by including the effect of direct $\sigma$-production.
We parametrize ${\cal F}$ as a sum of 
Breit-Wigner amplitudes for the relevant resonances 
(including $\sigma$-resonance)
and of the direct $2\pi$ production amplitude, following VMW-method.
This method is consistent with the requirement 
from $S$-matrix unitarity and from analyticity.
The VMW method had already
been applied\cite{rf6} to the $pp$-central collision, $pp\to pp \pi\pi$, 
the 
$J/\psi\to\omega\pi\pi$ decay and the $p\bar p\to 3\pi^0$ 
annihilation.\cite{rfppbar}
The large event concentration in low $m_{\pi\pi}$ region 
is explained by $\sigma$-production. 
Here we apply this method to the analyses of 
the hadronic decays of excited $\Upsilon$,
$\Upsilon (2S,3S)$, and $\psi (1S,2S)$.

\section{Method of the analyses}
We analyze the $m_{\pi\pi}$ spectra of the processes,
$\Upsilon (2S)\to\Upsilon (1S)\pi^+\pi^-$,\cite{rf7,rf8,rf9,rf10}
  $\Upsilon (3S)\to\Upsilon (1S)\pi^+\pi^-$,\cite{rf10} 
$\Upsilon (3S)\to\Upsilon (2S)\pi^+\pi^-$\cite{rf10} and 
$\psi (2S)\to J/\psi \pi^+\pi^-$\cite{rf11} following the VMW method in one-channel form.
The ${\cal F}$ is given by a coherent sum of $\sigma$-Breit-Wigner 
amplitude and of direct 
$2\pi$-production amplitude as
\begin{eqnarray}
{\cal F } &=& \frac{e^{i\theta_\sigma}r_\sigma }{m_\sigma^2 -s - i\sqrt{s}\Gamma_\sigma (s)}
+r_{2\pi}e^{i\theta_{2\pi}};\ \ \ 
\Gamma_\sigma (s) = \frac{g_\sigma^2 p_1(s)}{8\pi s}, \ \ 
p_1(s)=\sqrt{\frac{s}{4}-m_\pi^2} ,\nonumber
\end{eqnarray}
where $r_\sigma (r_{2\pi })$ is the $\sigma (2\pi )$ production coupling and 
$\theta_\sigma (\theta_{2\pi})$ corresponds to the production phase.
These parameters are process-dependent. 
The corresponding decay mechanism is depicted in 
Fig.~1(a), (b) for $\Upsilon$ decay and (c),(d) for $\psi (2S)$ decay.

\begin{figure}[t]
 \begin{center}
  \begin{tabular}[t]{c}
 \subfigure[ $\Upsilon' \to \Upsilon \sigma \to \Upsilon \pi \pi$ ]
    {\epsfxsize=4.4cm \epsffile{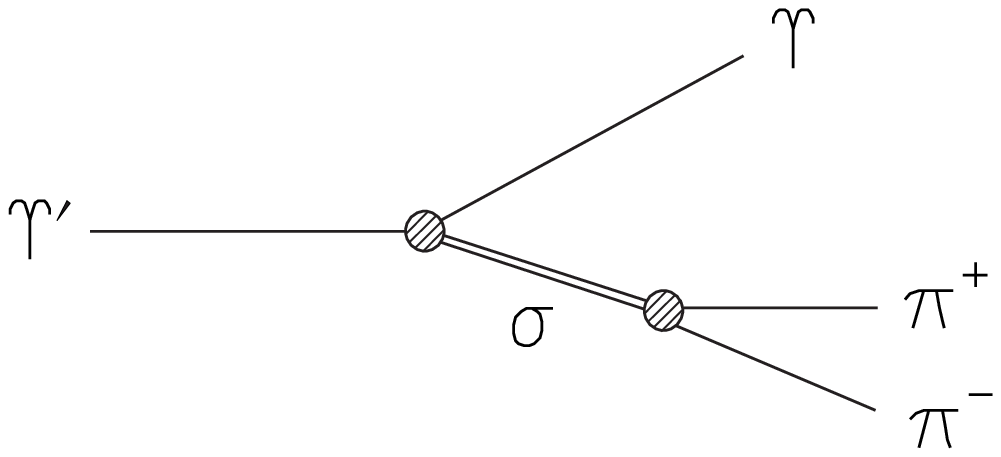}} 
\hspace{2em}
 \subfigure[ $\Upsilon' \to \Upsilon \pi \pi$ ]
    {\epsfxsize=4.4cm \epsffile{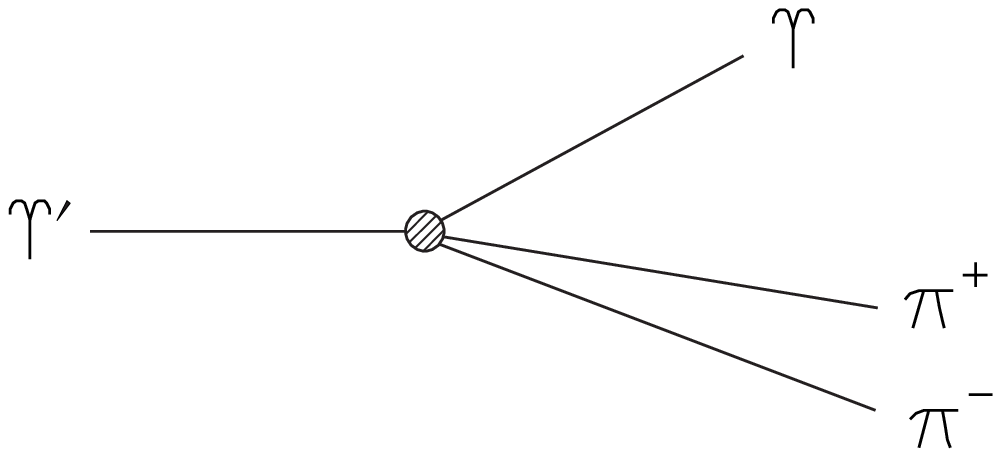}}     \vspace{-1em} \\
%
 \subfigure[$\psi(2S)$$\to$$J/ \psi$$\sigma$$\to$$J/ \psi$$\pi \pi$]
    {\epsfxsize=4.4cm \epsffile{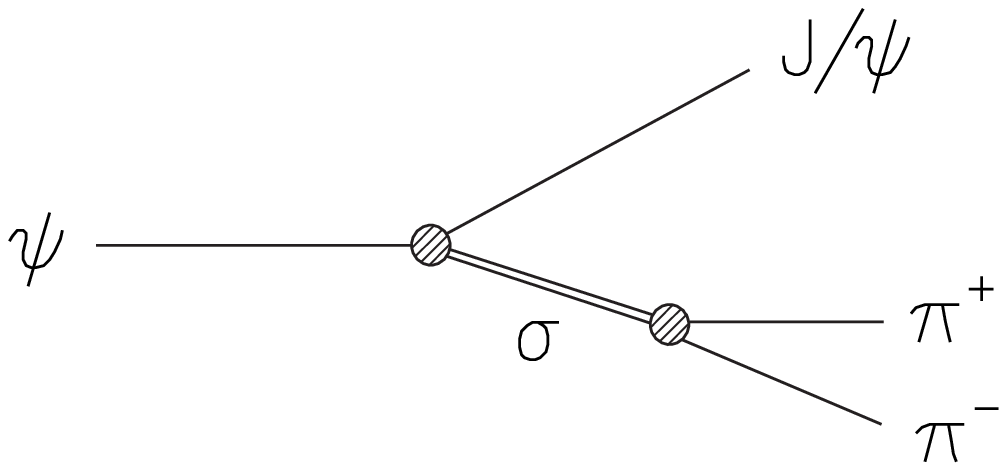}} 
\hspace{2em}
 \subfigure[ $\psi(2S) \to J/ \psi \pi \pi$ ]
    {\epsfxsize=4.4cm \epsffile{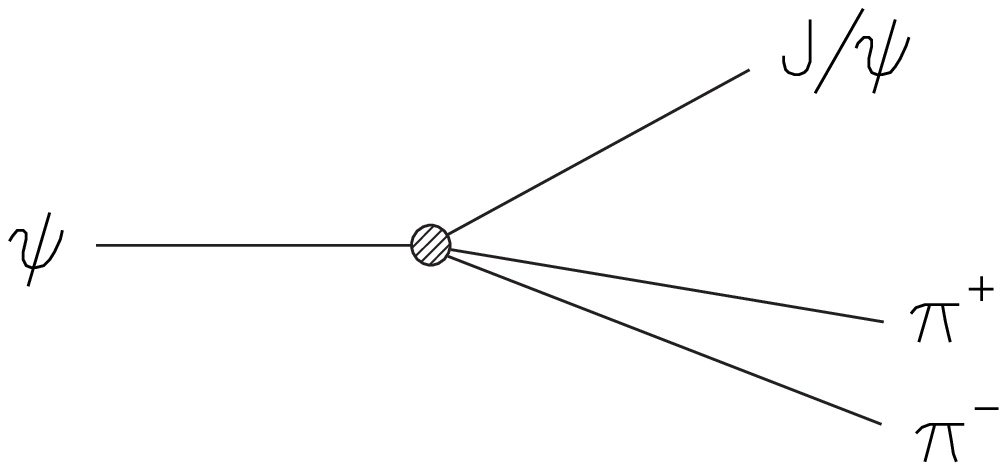}}     \vspace{-1em} \\
%
 \subfigure[$J/ \psi$$\to$$\phi$$\sigma$$\to$$\phi$$\pi \pi(K\overline{K})$ ]
    {\epsfxsize=4.4cm \epsffile{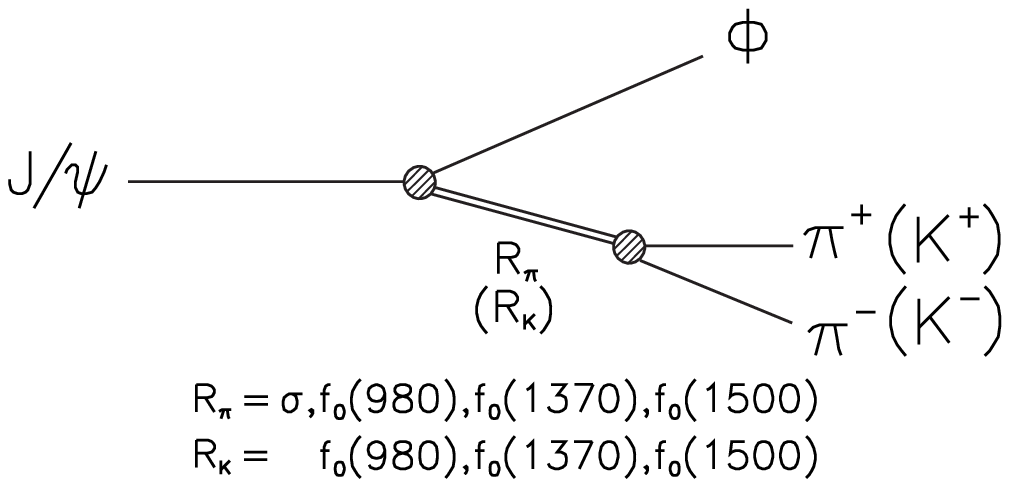}}  
\hspace{2em}
 \subfigure[$J/ \psi$$\to$$\phi$$\pi \pi$$(K\overline{K})$]
    {\epsfxsize=4.4cm \epsffile{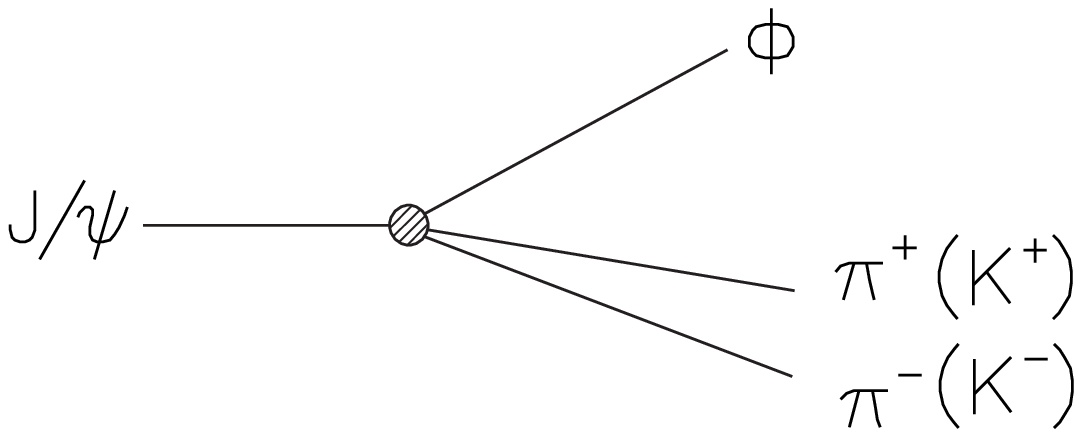}}      
  \end{tabular}
\caption{Mechanism of the relevant decays 
of the excited $\Upsilon (3S,2S)$, $\psi (2S)$ and the $J/\psi$. }
 \end{center}
\end{figure}
%
%

The $J/\psi\to \phi \pi^+\pi^- ,\ \phi K^+K^-$ process\cite{rf12} is analyzed by VMW-method in two-channel
form, where the $\pi\pi$ and $KK$ production amplitudes, ${\cal F}_{\pi\pi}$ and ${\cal F}_{KK}$,
respectively, are given by
\begin{eqnarray}
{\cal F}_{\pi\pi} &=& 
\frac{e^{i\theta_\sigma}r_\sigma }{m_\sigma^2 -s - i\sqrt{s}\Gamma_\sigma (s)}+\sum_{f_0} 
\frac{e^{i\theta_{f_0}}r_{f_0}g_{f_0\pi\pi} }{m_{f_0}^2 -s - i\sqrt{s}\Gamma_{f_0}^{\rm tot} (s)}
+r_{2\pi}e^{i\theta_{2\pi}};  \nonumber\\
{\cal F}_{KK} &=& \sum_{f_0} \frac{e^{i\theta_{f_0}}r_{f_0}g_{f_0KK} }{m_{f_0}^2 -s - i\sqrt{s}\Gamma_{f_0}^{\rm tot} (s)}
+r_{2K}e^{i\theta_{2K}};\ \ \ 
f_0=f_0(980),\ f_0(1370),\ f_0(1500)\nonumber\\
 & &  
\Gamma_{f_0}^{\rm tot} = \Gamma_{f_0}^{\pi\pi}+\Gamma_{f_0}^{KK}
         =\frac{p_1 g_{f_0\pi\pi}^2}{8\pi s} + \frac{p_2 g_{f_0KK}^2}{8\pi s}.
 \nonumber
\end{eqnarray}
The corresponding decay mechanism is shown in Fig.1 (e),(f).

By using the above formulas of production amplitudes 
the $m_{\pi\pi}$ or $m_{KK}$ spectra (that is, $\sqrt s$ spectra)
are given by
\begin{eqnarray} 
\frac{d \sigma}{d\sqrt s} &=& \frac{p(M'^2;M,\sqrt s)p_1(s)}{32\pi^3 M'^2}|{\cal F}|^2 ,\ \ \ 
p(M'^2;M,\sqrt s)=\frac{\sqrt{(M'^2-M^2-s)^2-4M^2 s}}{2\sqrt{s}}, \nonumber
\end{eqnarray}
where $M'(M)$ is the mass of the initial (final) quarkonium.

All the relevant spectra are fitted 
by using common values of the parameters, 
the mass of $\sigma$ $m_\sigma$ and 
the $\sigma\pi\pi$ coupling $g_{\sigma\pi\pi}$.  

\begin{figure}[t]
 \begin{center}
  \begin{tabular}[t]{c}
 \subfigure[ $\Upsilon(2S) \to \Upsilon(1S) \pi \pi$ ]
    {\epsfxsize=7cm \epsffile{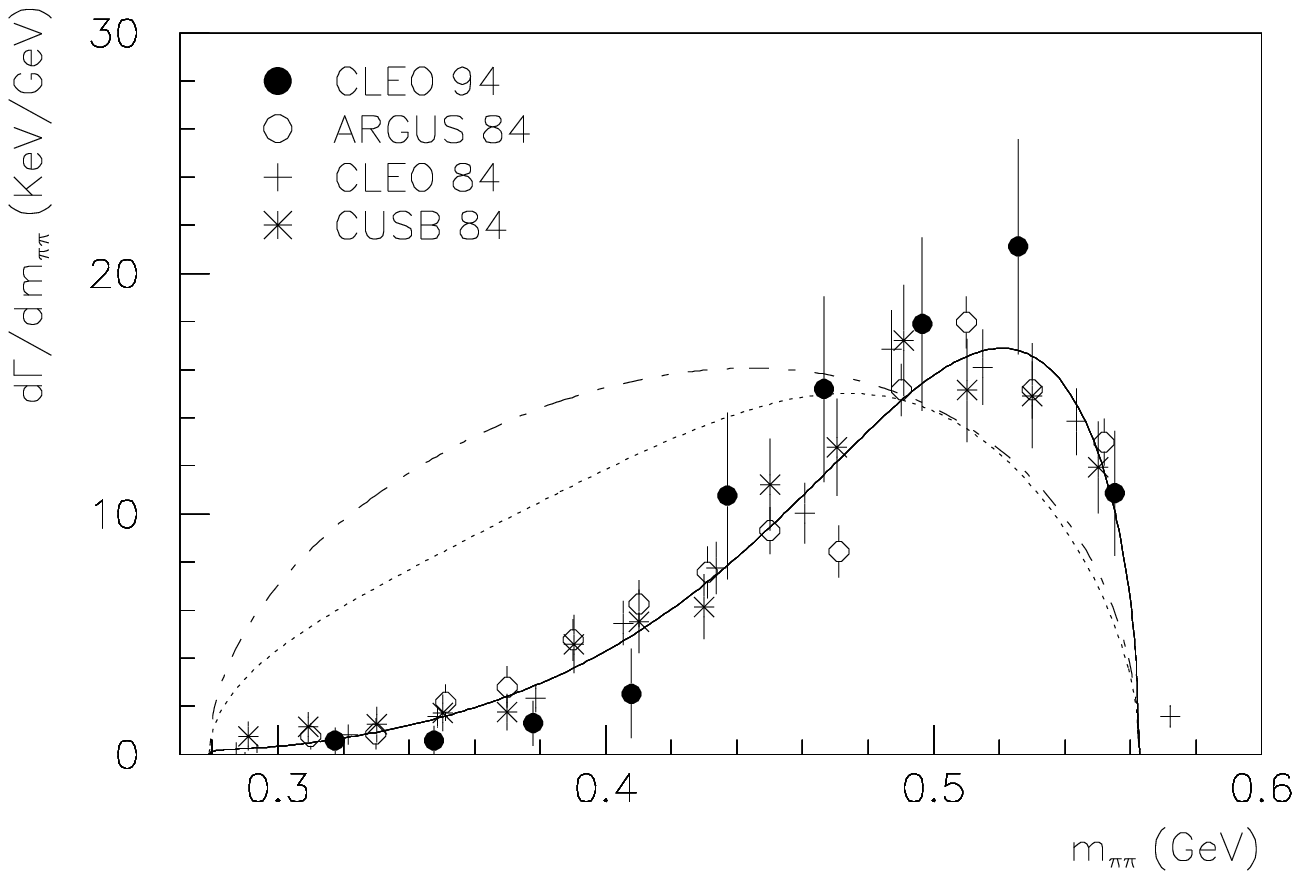}} 
\hspace{2em}
 \subfigure[ $\Upsilon(3S) \to \Upsilon(1S) \pi \pi$ ]
    {\epsfxsize=7cm \epsffile{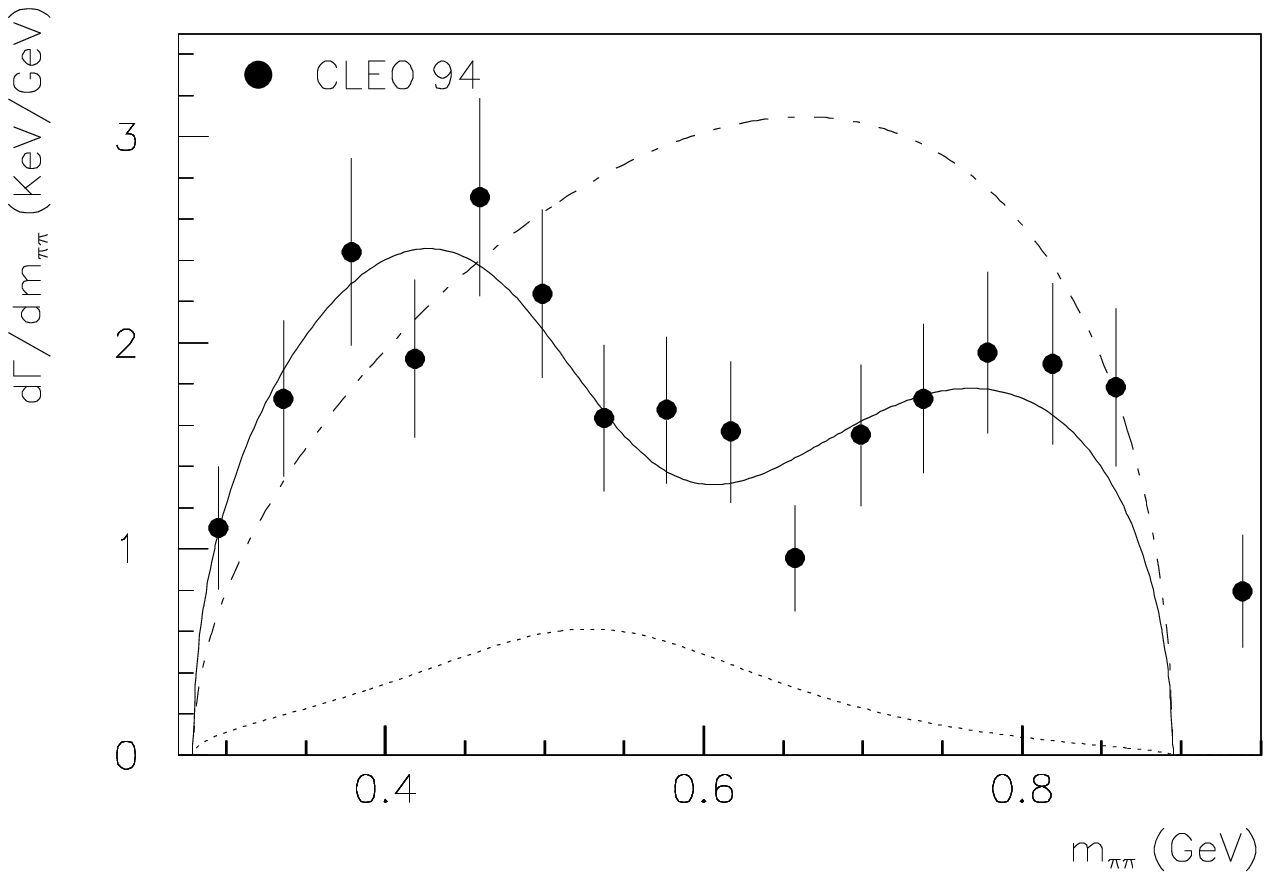}}     \vspace{-1em} \\
%
 \subfigure[$\Upsilon(3S) \to \Upsilon(2S) \pi \pi$ ]
    {\epsfxsize=7cm \epsffile{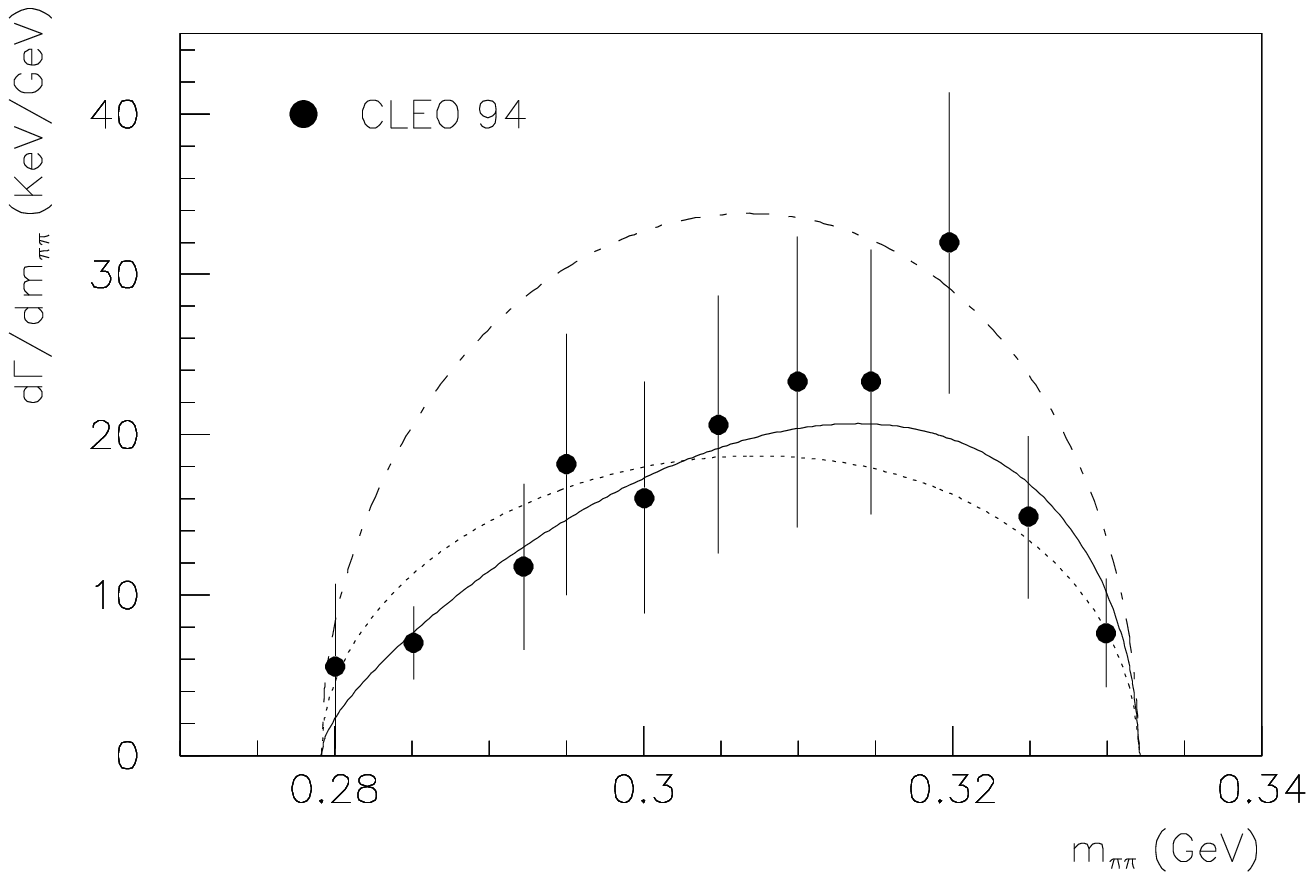}} 
\hspace{2em}
 \subfigure[ $\psi(2S) \to J/ \psi \pi \pi$ ]
    {\epsfxsize=7cm \epsffile{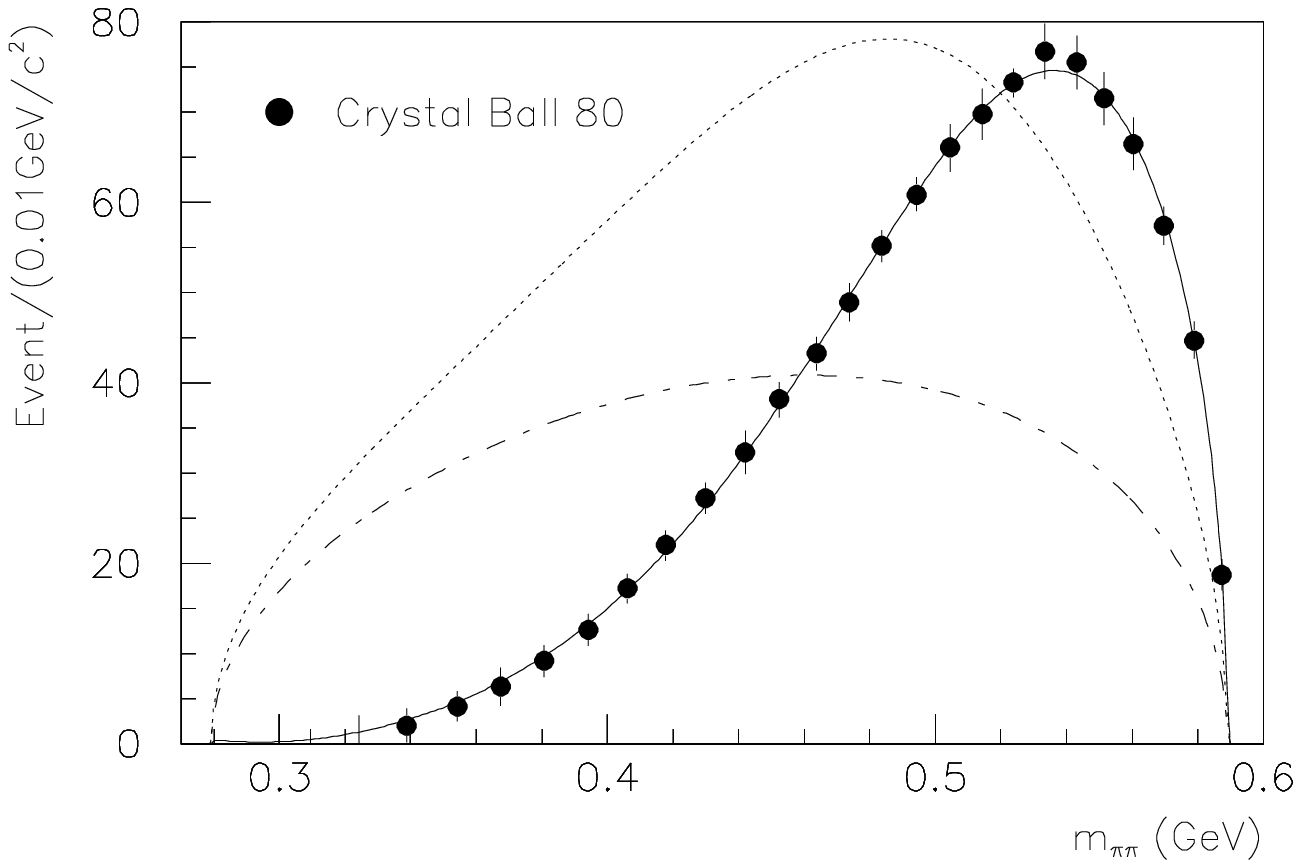}}     \vspace{-1em} \\
%
 \subfigure[ $J/ \psi \to \phi \pi \pi$ ]
    {\epsfxsize=7cm \epsffile{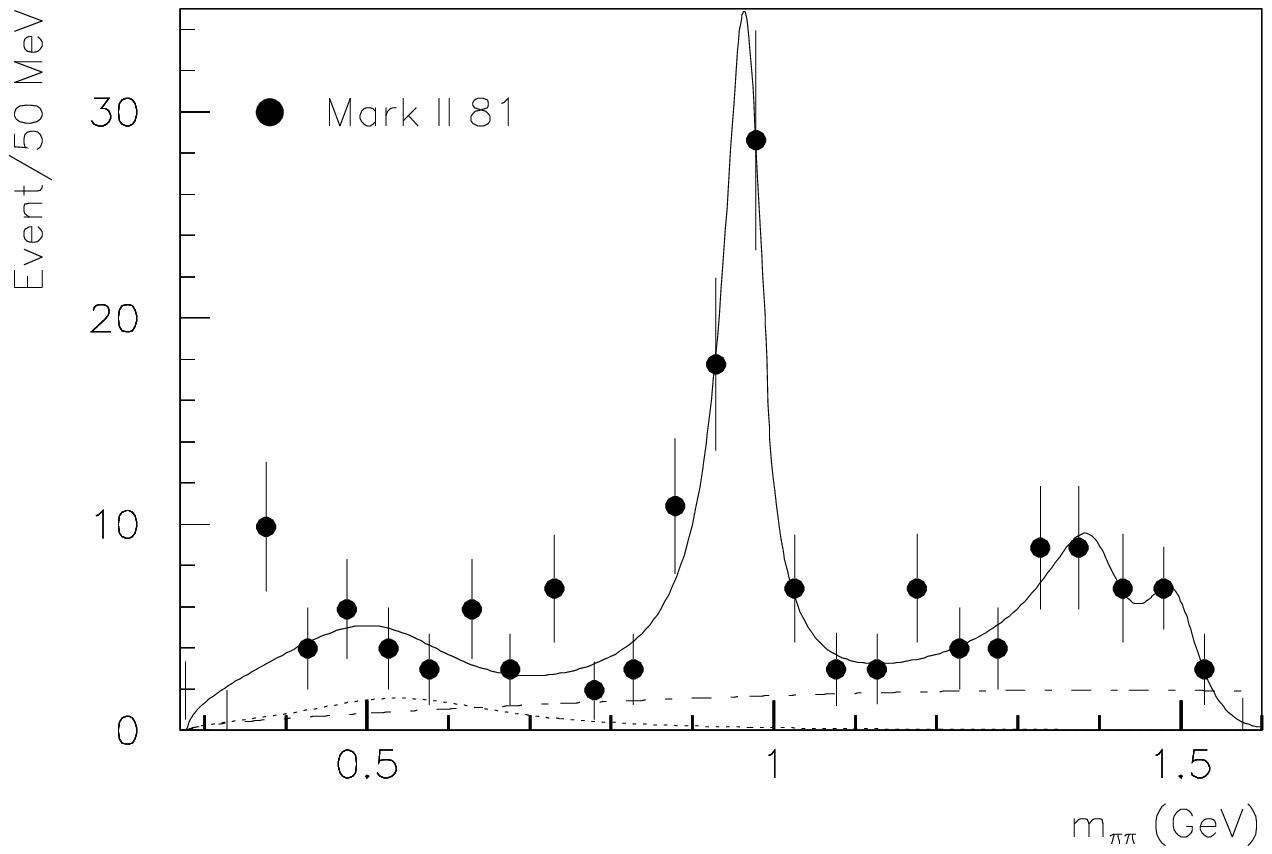}}  
\hspace{2em}
 \subfigure[ $J/ \psi \to \phi K K  $ ]
    {\epsfxsize=7cm \epsffile{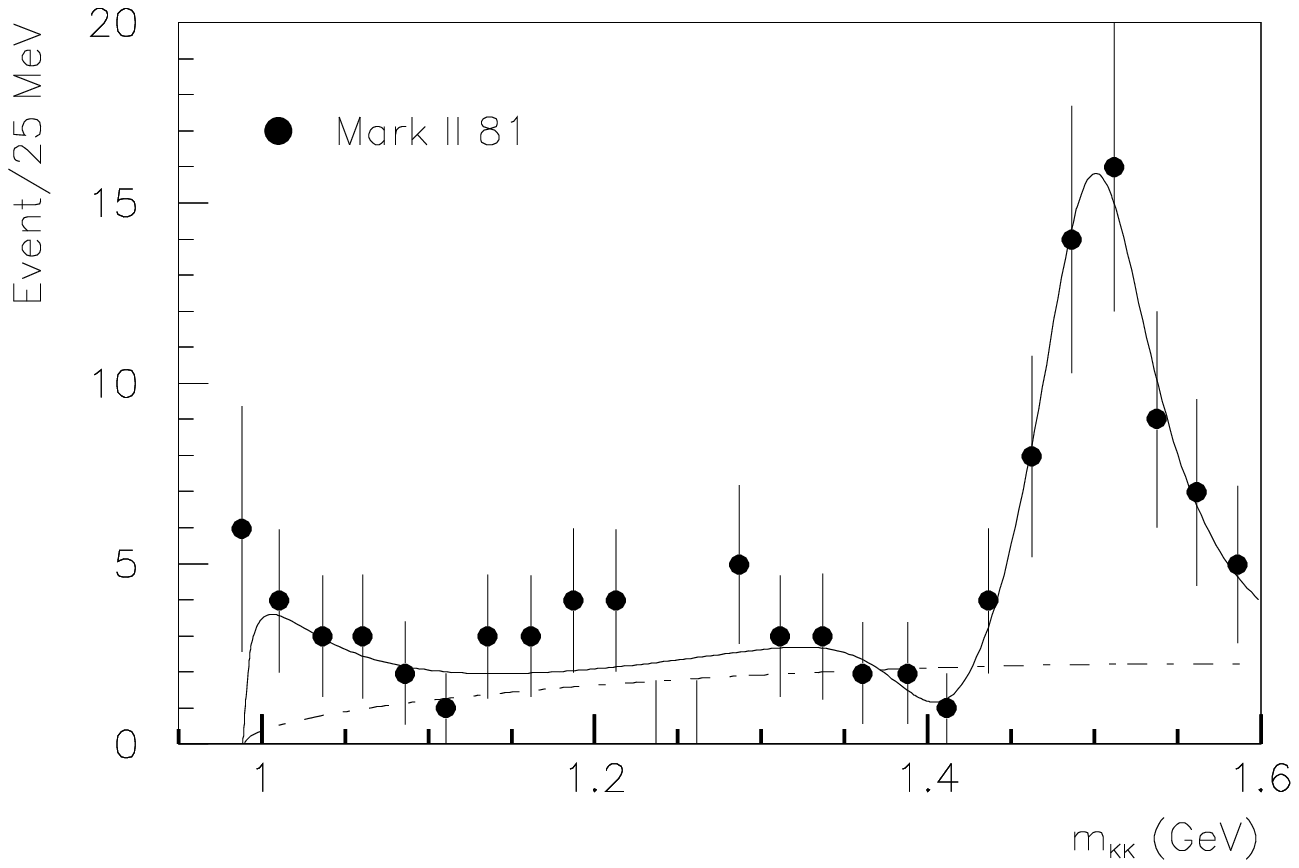}}      
  \end{tabular}
  \caption{
The result of the fit to the $\pi\pi$ (or $KK$) mass spectra of
(a) $\Upsilon (2S)\to \Upsilon (1S) \pi\pi$, 
(b) $\Upsilon (3S)\to \Upsilon (1S) \pi\pi$, 
(c) $\Upsilon (3S)\to \Upsilon (2S) \pi\pi$, 
(d) $\psi (2S)\to J/\psi  \pi\pi$, 
(e) $J/\psi \to \phi \pi\pi$, and    
(f)  $J/\psi \to \phi KK$ .  
The bold line represents the fit, while the dotted (dot-dashed) does the contribution
of $\sigma$(direct $2\pi$)-production. 
By multiplying appropriate proportional factors,
the different data sets are adjusted to the one 
with the largest number of events.
In (a) CLEO data\cite{rf7}, shown by black circle, with factor 1;
ARGUS\cite{rf8} by open circle with factor 1.5; CLEO\cite{rf9} by cross with factor 2;
CUSB\cite{rf10} by asterisk with factor 3.
Data in (b), (c) and (d)  are respectively, 
from CLEO\cite{rf7}, CLEO\cite{rf7} and Crystal Ball\cite{rf11}.
Data in (e) and (f) are from MARK II\cite{rf12}.
}
 \end{center}
\end{figure}

\begin{table}
\caption{Values of mass and width of resonances: 
$m_\sigma$ and $g_{\sigma\pi\pi}$ are taken to be 
common through all the processes.
The other resonance paramaters are relevant only 
in the $J/\psi\to\phi\pi\pi , \ \phi KK$ process.
Mass and coupling constant of $f_0(980)$ fall, respectively, 
on lower and upper limits in the fit.
 } 
\begin{center}
\begin{tabular}{|c|lll|ll|}
\hline
   & $m_\alpha$/MeV & $g_{\alpha\pi\pi}$/GeV & $\Gamma_{\alpha\pi\pi}$/MeV
                 & $g_{\alpha KK}$/GeV & $\Gamma_{\alpha KK}$/MeV  \\
\hline
$\sigma$ & 526 & 3.06 & 301 &  &   \\
\hline
$f_0(980)$ & 966 & 1.77 & 62 & 2.70  & 10 \\
$f_0(1370)$ & 1402 & 2.81 & 109 & 0.74  & 5 \\
$f_0(1500)$ & 1498 & 2.08 & 57 & 1.89  & 36 \\
\hline
\end{tabular}
\end{center}
\end{table}

\vspace{-0.5em} 
\begin{table}
\caption{Production couplings $r_\sigma$, $r_{2\pi}$ 
and phase $\theta_\sigma$
in one channel formula. 
$\theta_\sigma$ is set to be zero and
only the relative ratios of $r$ are meaningful.  
 } 
\begin{center}
\begin{tabular}{|c|l|l|l|l|}
\hline
   & $\Upsilon (2S\to 1S)$ & $\Upsilon (3S\to 1S)$ & $\Upsilon (3S\to 2S)$
                 & $\psi (2S\to 1S)$ \\
\hline
$r_\sigma$ & 6031 & 103 & 3138 & 659 \\

$r_{2\pi}$ &  38512 & 1381 & 21270 & 2986 \\
$\theta_{2\pi}$ & 197 & 322 & 154 & 199 \\
\hline
\end{tabular}
\end{center}
\end{table}
 
\vspace{-0.5em} 
\begin{table}
\caption{Production couplings $r_\sigma$, $r_{2\pi}$ and 
phase $\theta_\sigma$ in two channel formula
of the decay $J/\psi\to\phi\pi\pi ,\ \phi KK$. 
$\theta_\sigma$ is set to be zero and  
only the relative ratios of $r$ are meaningful.  
 } 
\begin{center}
\begin{tabular}{|c|l|l|l|l|l|l|}
\hline
   & $\sigma$ & $f_0(980)$ & $f_0(1370)$ & $f_0(1500)$ & $2\pi$ & $2K$ \\
\hline
$r$ & 41GeV$^2$ & 42GeV & 127GeV & 78GeV & 195 & 238 \\
$\theta$(degree) & 0(fixed) & 355 & 104 & 329 & 382 & 56 \\
\hline
\end{tabular}
\end{center}
\end{table}

\begin{table}
\caption{$\chi^2$ values for the respective data sets. Total $\chi^2$ is
$\chi^2 /(N_{\rm data}-N_{\rm param})=86.5/(150-37)$.
 } 
\begin{center}
\begin{tabular}{|c|l|l|l|l|}
\hline
   & $\Upsilon (2S\to 1S)$ & $\Upsilon (2S\to 1S)$ 
    & $\Upsilon (2S\to 1S)$ & $\Upsilon (2S\to 1S)$ \\
   & CLEO\cite{rf7}   & CLEO\cite{rf8}  & CLEO\cite{rf9}  & CLEO\cite{rf10} \\
\hline
$N_{\rm data}$ & 14 & 10 & 14 & 10 \\
$\chi^2$ & 27.0 & 4.8 & 6.8 & 8.9 \\
\hline
\hline
    & $\Upsilon (3S\to 1S)$ & $\Upsilon (3S\to 2S)$ 
    & $\psi (2S\to 1S)$ & $J/\psi \to \phi\pi\pi ,\ \phi KK$ \\
   & CLEO\cite{rf7}   & CLEO\cite{rf7}  & Crystal Ball\cite{rf11}  & MARK II\cite{rf12} \\
\hline
$N_{\rm data}$ & 15 & 11 & 26 & 26+24 \\
$\chi^2$ & 9.8 & 3.4 & 3.3 & 16.7+16.7 \\
\hline
\end{tabular}
\end{center}
\end{table}

\section{Results and Conclusion}
The results of our analysis are shown in Fig.~2. 
The spectra of five different processes 
((a) to (e) in Fig. 1) are reproduced well 
by the interference between the $\sigma$ Breit-Wigner
amplitude with 
\begin{eqnarray}  
m_\sigma &=& 526 ^{+48}_{-37} \ \ \ {\rm MeV}, \ \ \ 
\Gamma_\sigma =301^{+145}_{-100} \ \ {\rm MeV}
\label{eq3}
\end{eqnarray}
and the direct $2\pi$ production amplitude.
The total $\chi^2$ is $\chi^2/(N_{\rm data}-N_{\rm
param})=86.5/(150-37)=0.77$.
The contribution of $\sigma$ and of direct $2\pi$ production amplitudes to the spectra
are given, respectively, by dot and dot-dashed lines in this figure. 
It is notable that 
the destructive interference between $\sigma$ amplitude and $2\pi$ amplitudes 
explains the suppression of the spectra 
in the threshold region of $\Upsilon (2S\to 1S)$ and $\psi (2S\to 1S)$ decays,
while in $\Upsilon (3S\to 1S)$ decay these two amplitudes interfere constructively, 
and the steep increase from the threshold is reproduced.
These threshold behaviors of the production amplitudes are\cite{rf4} 
shown to be consistent with the restriction from chiral symmetry.
The double peak structure is also reproduced well by the above 
interference, that is, the destructive interference around the 
$\sigma$-peak position.

The obtained values of parameters are given in 
Table I for masses and widths of resonances and in Table
II(III) for production couplings and phases in one channel 
(two channel) formula.
The $\chi^2$ values for respective data sets are given in Table IV.

The property of $\sigma$, given in Eq.(\ref{eq3}) is almost consistent with the results
obtained in our previous $\pi\pi$ phase shift analysis,\cite{rf3}
$m_\sigma$=535$\sim$675MeV and
$\Gamma_\sigma$=385$\pm$70MeV. 
These results give strong evidence for
existence of the light $\sigma (500$--$600)$.

\end{document}